\documentclass[twocolumn,showpacs,preprintnumbers,amsmath,amssymb]{revtex4}
\usepackage{graphicx}
\usepackage{dcolumn}
\usepackage{bm}
\begin{document}
\title{Conductivity of a spin-polarized two-dimensional electron liquid in the ballistic regime}
\author{A.~A. Shashkin, E.~V. Deviatov, V.~T. Dolgopolov, and A.~A. Kapustin}
\affiliation{Institute of Solid State Physics, Chernogolovka, Moscow District 142432, Russia}
\author{S. Anissimova, A. Venkatesan, and S.~V. Kravchenko}
\affiliation{Physics Department, Northeastern University, Boston, Massachusetts 02115, U.S.A.}
\author{T.~M. Klapwijk}
\affiliation{Kavli Institute of Nanoscience, Delft University of Technology, 2628 CJ Delft, The Netherlands}
\begin{abstract}
In the ballistic regime, the metallic temperature dependence of the
conductivity in a two-dimensional electron system in silicon is found
to change non-monotonically with the degree of spin polarization. In
particular, it fades away just before the onset of complete spin
polarization but reappears again in the fully spin-polarized state,
being, however, suppressed relative to the zero-field case. Analysis
of the degree of the suppression allows one to distinguish between
the screening and the interaction-based theories.
\end{abstract}
\maketitle

\section{INTRODUCTION}

Much interest has been attracted recently to the anomalous properties
of low-disordered, strongly correlated two-dimensional (2D) electron
systems. The effects of electron-electron interactions are especially
strong in silicon metal-oxide-semiconductor field-effect transistors
(MOSFETs) (for recent reviews, see
Refs.~\cite{kravchenko04,shashkin05}), but they are also pronounced
in other systems like GaAs/AlGaAs \cite{tan05} and Si/SiGe
heterostructures \cite{lai05}. Interactions lead, in particular, to
critical behavior of the Pauli spin susceptibility
\cite{shashkin01,vitkalov01} and sharply increasing effective mass at
low electron densities \cite{shashkin02,shashkin03,anissimova06}.
These phenomena (at least in Si MOSFETs) are not dominated by spin
exchange effects, since the Land\'e $g$ factor is found to be close
to its value in a bulk semiconductor and the effective mass is
insensitive to the degree of spin polarization. At the same time,
spin effects are the origin of the strong positive magnetoresistance
in parallel magnetic fields (see, e.g.,
Refs.~\cite{simonian97,yoon00}). Therefore, one can probe the spin
effects by studying peculiarities of a spin-polarized 2D electron
system. The case of complete spin polarization of the electron system
is especially interesting because it is the simplest from the
theoretical point of view.

It is known that application of a parallel magnetic field causes
giant (orders of magnitude) positive magnetoresistance and fully
suppresses the metallic state near the 2D metal-insulator transition
\cite{simonian97,yoon00,dolgopolov92}. However, if the electron
density is not too low (ballistic regime, $k_BT\gtrsim\hbar/\tau$
\cite{zala01}), the metallic temperature dependence of conductivity
has been found to persist in the fully spin-polarized state
\cite{okamoto00,mertes01,tsui05}. (Note that in silicon-based devices
studied in these papers, electrons possess the valley degree of
freedom, which survives in the fully spin-polarized state.)
Conductivity of silicon MOSFETs in this regime was studied in
Ref.~\cite{tsui05}. However, for much of their data (particularly at
relatively high temperatures and/or electron densities), the complete
spin polarization was in fact not reached as a result of the
insufficiently high magnetic fields used.

Theoretically, linear-in-temperature corrections to the zero-field
conductivity in the ballistic regime were calculated in
Ref.~\cite{gold86}. In the newer theory \cite{zala01}, the exchange
interaction terms were treated more carefully. However, it turned out
that at $B=0$, both the screening \cite{gold86} and the
interaction-based \cite{zala01} theories describe the
temperature-dependent conductivity equally well \cite{shashkin04a}.
To distinguish between them, studies of the effect of the parallel
magnetic field on conductivity may be helpful.

Here we experimentally study the transport properties of a 2D
electron system in silicon in parallel magnetic fields at different
degrees of spin polarization in the ballistic regime. We show that in
a completely spin-polarized state, disorder effects are dominant when
approaching the regime of strong localization, which is in contrast
to the behavior of the unpolarized state in low-disordered 2D
electron systems. The temperature-dependent correction to the elastic
relaxation time is found to change strongly with the degree of spin
polarization, reaching a minimum just below the onset of full spin
polarization, where the conductivity is practically independent of
temperature. In the fully spin-polarized state, the correction
mentioned above is about two times weaker than that in $B=0$ at the
same electron density. This is consistent with what one expects
according to the simple version of screening theory
\cite{dolgopolov00}.

\section{EXPERIMENTAL TECHNIQUE AND SAMPLES}

Measurements were made in an Oxford dilution refrigerator on
(100)-silicon MOSFETs with peak electron mobilities of about
3~m$^2$/Vs at 0.1~K. The resistance was measured with a standard
4-terminal technique at a low frequency (1~Hz) to minimize the
out-of-phase signal. Excitation current was kept low enough (below
1~nA) to ensure that measurements were taken in the linear regime of
response. Contact resistances in these samples were minimized by
using a split-gate technique that allows one to maintain a high
electron density in the vicinity of the contacts (about $1.5\times
10^{12}$~cm$^{-2}$), regardless of its value in the main part of the
sample.

\begin{figure}
\scalebox{0.42}{\includegraphics[clip]{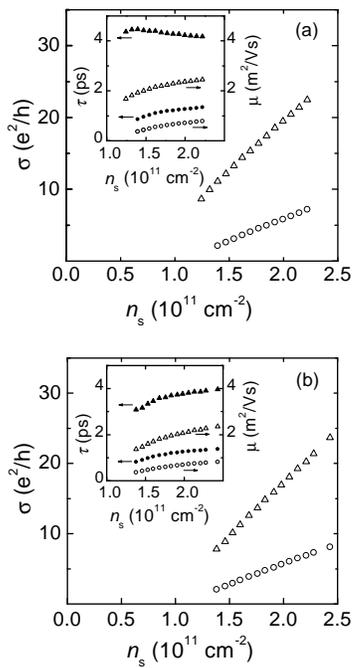}}
\caption{\label{fig1} Conductivity, mobility, and elastic scattering
time vs.\ electron density at a temperature of 0.1~K for
spin-unpolarized (triangles) and fully spin-polarized (circles)
states in two slightly different samples A at $B=0$ and 9.5~T (a) and
B at $B=0$ and 14~T (b).}
\end{figure}

In Fig.~\ref{fig1}, we show low-temperature conductivity as a
function of electron density, $n_s$, in the metallic regime (i.e.,
when the conductivity $\sigma>e^2/h$) for fully spin-polarized and
spin-unpolarized states in two slightly different samples. As the
electron density is decreased, the conductivity significantly drops
as a result of a decrease in the electron mobility $\mu$ (insets to
Fig.~\ref{fig1}). This mobility decrease originates from the effects
of both disorder and electron-electron interactions. The former
determines the elastic relaxation time, $\tau$, while the latter is
responsible for the enhanced effective mass
\cite{shashkin02,shashkin03,anissimova06}. Taking the effective mass
values from Ref.~\cite{shashkin03}, we have calculated $\tau$ as
shown in the insets to Fig.~\ref{fig1}. For the spin-unpolarized
state, the behavior of $\tau$ for the two samples is qualitatively
different, although the mobilities are very similar. In sample A, the
decrease in $\mu$ at low densities is dominated by electron-electron
interactions (increasing effective mass), while in sample B, disorder
effects are more pronounced and lead to $\tau$ decreasing at low
$n_s$. In the fully spin-polarized state, the disorder effects
prevail in both samples.

\section{RESULTS}

\begin{figure}
\scalebox{0.42}{\includegraphics[clip]{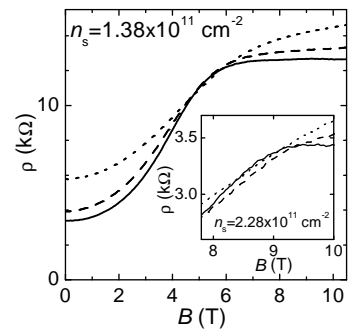}}
\caption{\label{fig2} Magnetoresistance at temperatures 0.5 (solid
line), 0.8 (dashed line), and 1.2~K (dotted line) on sample B. The
inset shows a detailed view of the magnetoresistance just before the
onset of complete spin polarization.}
\end{figure}

\begin{figure}
\scalebox{0.42}{\includegraphics[clip]{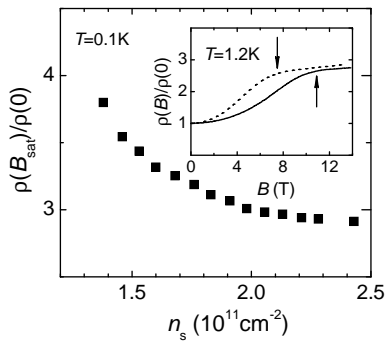}}
\caption{\label{fig3} Change of the resistance ratio,
$\rho(B_{\text{sat}})/\rho(0)$, with electron density in sample B.
The inset shows the normalized magnetoresistance, measured at the
highest temperature used in this experiment, at electron densities
$1.46\times10^{11}$~cm$^{-2}$ (dotted line) and
$2.13\times10^{11}$~cm$^{-2}$ (solid line). The field of resistance
saturation is marked by arrows.}
\end{figure}

Experimental traces of the parallel-field magnetoresistance at
different temperatures are displayed in Fig.~\ref{fig2}. The
low-temperature resistivity, $\rho$, rises with $B$ and saturates
above a certain $n_s$-dependent magnetic field, $B_{\text{sat}}(0)$,
corresponding to the onset of complete spin polarization of the 2D
electrons \cite{okamoto99}. Increasing the temperature leads to
smearing the dependences so that the resistance saturation occurs at
higher magnetic fields. In other words, the saturation field
increases as the temperature is increased \cite{vitkalov01}. The
resistivity rises appreciably with increasing temperature in both
$B=0$ and $B>B_{\text{sat}}$, here the saturation field
$B_{\text{sat}}$ corresponds to the highest temperature used in the
experiment. The magnetoresistance at two electron densities measured
at the highest temperature used, $T\approx1.2$~K, is shown in the
inset to Fig.~\ref{fig3}. The fact that the magnetoresistance
saturates at sufficiently high magnetic fields confirms that the full
spin polarization is reached in our experiment even at this
temperature. As seen from Fig.~\ref{fig2}, just below
$B_{\text{sat}}(0)$ the resistivity practically does not depend on
temperature up to the highest temperatures used. The validity of this
effect, which has also been observed in Refs.~\cite{mertes01,tsui05},
has been verified at ten electron densities in the range between
$1.38\times10^{11}$ and $2.42\times10^{11}$~cm$^{-2}$. We would like
to emphasize that the flattening of $\sigma(T)$ just below the onset
of complete spin polarization makes it difficult to analyze the data
for $\sigma(T)$ obtained in a fixed magnetic field or in a narrow
field region \cite{mertes01,tsui05} as the complete spin polarization
may have not been reached at higher temperatures and/or electron
densities.

The low-temperature ratio $\rho(B_{\text{sat}})/\rho(0)$ vs.\
electron density is shown in Fig.~\ref{fig3}. In agreement with the
previously obtained data, it increases weakly with decreasing $n_s$,
being close to the value $\rho(B_{\text{sat}})/\rho(0)=4$ predicted
by the theory of the spin-polarization-dependent screening of a
random potential \cite{dolgopolov00}. As seen from the inset, the
ratio $\rho(B_{\text{sat}})/\rho(0)$ diminishes somewhat at higher
temperatures.

The normalized conductivity as a function of temperature in fully
spin-polarized and spin-unpolarized states is depicted in the inset
to Fig.~\ref{fig4}. The correction to $\sigma/\sigma(0)$ is linear in
temperature with the slope given by $A^*=-\sigma_0^{-1}d\sigma/dT$,
where $\sigma_0=\sigma(0)$ is obtained by linear extrapolation of the
data to $T=0$. We emphasize that the so-defined slopes do not depend
on $\tau$ and are therefore different from the slopes defined in
Ref.~\cite{tsui05} as $d\sigma/dT$: the ratio of the slopes
$r\equiv[d\sigma(B_{\text{sat}})/dT]/[d\sigma(0)/dT]$ used there is
smaller by a factor of $\tau(0)/\tau(B_{\text{sat}})\approx4$ than
the ratio $A^*(B_{\text{sat}})/A^*(0)$ used in this paper.

In Fig.~\ref{fig4}, we show how the ratio of the slopes
$A^*(B_{\text{sat}})/A^*(0)$ for completely spin-polarized and
unpolarized states changes with electron density. Being approximately
equal to 0.5 at low $n_s$, the slope ratio increases weakly with
increasing $n_s$ but remains less than one in the range of electron
densities studied. Thus, the metallic behavior of the normalized
conductivity is always suppressed in the fully spin-polarized state.

\section{DISCUSSION}

We give a qualitative account of the absence of the $\sigma(T)$
dependence just below the onset of complete spin polarization. In the
magnetic field $B=B_{\text{sat}}(0)$, the degree of spin polarization
decreases linearly with temperature: $\xi=1-\gamma k_BT/E_F$ (where
the factor $\gamma\sim1$ and $E_F$ is the Fermi energy of the
spin-polarized 2D electrons). The increase in the number of electrons
with opposite spin direction naturally leads to increasing
conductivity. Therefore, near the onset of complete spin polarization
there exists another contribution to the temperature-dependent
conductivity, whose sign is opposite compared to the conventional
screening behavior of $\sigma(T)$. In the simple version of the
screening theory \cite{dolgopolov00}, the derivative $d\rho/d\xi$ at
$T=0$ tends to infinity as one approaches the field
$B_{\text{sat}}(0)$ from below. This feature will obviously be
smeared out at finite temperatures and/or due to the disorder present
in real electron systems. It is clear that depending on disorder
strength, two opposite contributions to the linear-in-$T$ correction
to conductivity can in principle balance each other \cite{rem}.

It is worth comparing the behavior of 2D electron system in Si
MOSFETs to that in another two-valley system, Si/SiGe quantum wells.
Transport properties of the latter system have been found to be very
similar to those of silicon MOSFETs
\cite{lai05,okamoto00,dolgopolov03}, although the disordered
potential in both cases is different resulting, particularly, from
the presence/absence of a spacer. However, the peculiarities near the
onset of complete spin polarization are less pronounced in Si/SiGe
quantum wells than in MOSFETs: only a weakening, but not absence, of
the temperature dependence of the resistance has been observed in the
metallic regime in a partially spin-polarized state \cite{lai05}.
Theoretically, the effect of the weakening of the $\sigma(T)$
dependence near the onset of complete spin polarization has been
found for the 2D electrons in Si/SiGe quantum wells in the frames of
screening approach \cite{hwang05}.

\begin{figure}
\scalebox{0.42}{\includegraphics[clip]{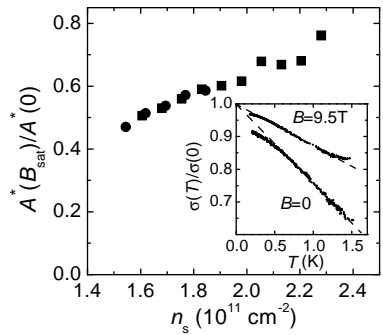}}
\caption{\label{fig4} The slope ratio as a function of electron
density for samples A (circles) and B (squares). The inset shows the
temperature dependence of the conductivity in both the fully
spin-polarized state for $n_s=1.85\times10^{11}$~cm$^{-2}$ and the
unpolarized state for $n_s=1.7\times10^{11}$~cm$^{-2}$ on sample A.
The dashed lines are fits of the linear interval of the dependence.}
\end{figure}

We now compare the experimental ratio of the slopes
$A^*(B_{\text{sat}})/A^*(0)$ with theoretical predictions. As we have
already mentioned, in zero magnetic field, both the
temperature-dependent screening theory \cite{gold86} and the
interaction-based theory \cite{zala01} describe reasonably well the
available experimental data for $\sigma(T)$ in silicon MOSFETs
\cite{shashkin05,shashkin04a,sarma04}. For the fully spin-polarized
state, however, their predictions are very different. In theory
\cite{zala01}, the ratio $A^*(B_{\text{sat}})/A^*(0)$ (for a
two-valley 2D system) is formally equal to $(1+4F_0^\sigma)/(1+\alpha
F_0^\sigma)$ \cite{remark}, once the effective mass, as well as the
$g$ factor, are independent of the degree of spin polarization
\cite{kravchenko04,shashkin05}. Here the interaction parameter
$F_0^\sigma$ is responsible for the renormalization of the $g$ factor
through $g=2/(1+F_0^\sigma)$, the coefficient $\alpha=8$ if
$T<\Delta_v$ and $\alpha=16$ if $T>\Delta_v$, and $\Delta_v$ is the
valley splitting. For negative $F_0^\sigma$, the observed slope ratio
(Fig.~\ref{fig4}) cannot at all be attained within the approach
\cite{zala01}: based on the $B=0$ data for $F_0^\sigma$
\cite{shashkin02}, considerably smaller values of the slope ratio are
expected compared to the experiment. On the other hand, according to
the screening theory in its simple form (ignoring the local field
corrections), the ratio $A^*(B_{\text{sat}})/A^*(0)$ is equal to 0.5,
as inferred from doubling the Fermi energy due to the lifting of the
spin degeneracy \cite{dolgopolov00}. This value is close to the
experimental finding. The observed decrease of the slope ratio at low
electron densities is likely to be similar to the behavior of the
resistance ratio mentioned above (see Fig.~\ref{fig3}). Concerning
the data for Si/SiGe quantum wells \cite{lai05}, one can evaluate the
slope ratio for $n_s=0.515\times10^{11}$~cm$^{-2}$ at about 0.45,
which is consistent with our results.

\section{CONCLUSION}

In summary, we have found that in the ballistic regime, the metallic
temperature dependence of the conductivity in a two-dimensional
electron system in silicon changes non-monotonically with the degree
of spin polarization. It fades away just below the onset of complete
spin polarization but reappears again, being suppressed, in the fully
spin-polarized state. A qualitative account of the effect of the
disappearance of the $\sigma(T)$ dependence near the onset of
complete spin polarization is given. While in zero magnetic field
both the temperature-dependent screening theory and the
interaction-based theory provide a reasonably good description of
experimental data for the temperature-dependent conductivity in the
ballistic regime, the results obtained in the fully spin-polarized
state favor the screening theory in its simple form.

\acknowledgments

We gratefully acknowledge discussions with I.~L. Aleiner, A. Gold,
and V.~S. Khrapai. This work was supported by the RFBR, the Programme
``The State Support of Leading Scientific Schools'', the National
Science Foundation grant DMR-0403026, and the ACS Petroleum Research
Fund grant 41867-AC10. EVD acknowledges the Russian Science Support
Foundation.

\end{document}